\documentclass[showpacs,aps,twocolumn]{revtex4}
\usepackage{epsfig,graphicx,float,pst-all,rotating}

\begin{document}

\title{Critical point symmetries in boson-fermion systems. The case of
  shape transition in odd nuclei in a multi-orbit model} 

\author{C.E. Alonso$^1$, J.M. Arias$^1$, and A. Vitturi$^{1,2}$ \\}

\affiliation{$^1$ Departamento de F\'{\i}sica At\'omica, Molecular y
  Nuclear, Facultad de F\'{\i}sica \\ Universidad de Sevilla, Apartado~1065, 
41080 Sevilla, Spain \\
$^2$ Dipartimento di Fisica Galileo Galilei and INFN,
Via Marzolo 8, 35131 Padova, Italy}

\begin{abstract}
We investigate phase transitions in boson-fermion systems. We propose
an analytically solvable model (E(5/12)) to describe odd nuclei 
at the critical point in the transition from the spherical  to
$\gamma$-unstable behaviour.  In the model, 
a boson core described within the Bohr Hamiltonian interacts with an unpaired particle 
assumed to be moving in the three single particle orbitals
j=1/2,3/2,5/2. Energy spectra and electromagnetic transitions at the  
critical point  compare well with the results obtained within the
Interacting Boson Fermion Model, with a boson-fermion Hamiltonian  
that describes the same physical situation.

\pacs{21.60.-n, 21.60.Fw, 21.60.Ev}  
\end{abstract}

\maketitle
 
The occurrence of shape transitional behaviour characterizes both
even-even and odd nuclei. Because of the further complexity embodied
in the description of the coupling of the odd particle to the even
core, formalisms describing phase transitions have been initially
mainly devised for even systems. Only recently the study of phase
transitions in odd-nuclei has been revitalized. The first case
considered by Iachello\cite{E54,cam,caprio} has been the case of a j=3/2
fermion coupled to a boson core that undergoes a transition from
spherical to $\gamma$-unstable. At the critical point, an elegant
analytic solution, called E(5/4), has been obtained starting from the
Bohr Hamiltonian. The possible occurrence of this symmetry in
$^{135}$Ba has been recently suggested\cite{135ba}. However, the
restriction of the fermion space to a single-j orbit makes difficult
the identification of a particular nucleus as critical. 

In this letter we consider the richer case of a collective core
coupled to a particle moving in the j=1/2, 3/2 and 5/2 single-particle
orbits, with the boson part undergoing a transition from sphericity to
deformed $\gamma$-instability.  This situation is likely to occur in
the mass region around A=190, for example in odd-neutron Pt where the
relevant active orbitals around the Fermi surface are  the
$3p_{1/2},3p_{3/2}$ and $2f_{5/2}$, in odd-proton Ir isotopes with
active orbits $2d_{5/2}, 2d_{3/2}$ and $3s_{1/2}$, and also in the
region around A=130, in odd-neutron Ba isotopes with relevant orbits
$2d_{5/2}, 2d_{3/2}$ and $3s_{1/2}$, just to cite a few of them. At the
critical point we extend the E(5/4) model to the multi-j case,
obtaining a new analytic solution, which we will call E(5/12). Knowing
the analytic form of the eigenfunctions, it is then easy to calculate
spectroscopic observables, such as electromagnetic transition
probabilities.  To check the validity of the model, we will also
present more microscopic calculations based on the Interacting Boson
Fermion Model\cite{IVI91}, with a choice of the 
boson-fermion Hamiltonian that describes the same physical situation.
We will show that the E(5/12) and IBFM models give comparable
behaviours.  

We will then consider the coupling of a collective core, described
within the quadrupole collective Bohr Hamiltonian, with an odd
particle moving  
in the 1/2,3/2,5/2 orbitals. One can use the separation of the
single-particle angular momentum into a pseudo-spin and pseudo-orbital
angular momentum (0 and 2) and assume no coupling due to the
pseudo-spin. 
The pseudo-orbital part transforms as the representations
($\tau_1^F,0$) of $O^F(5)$, with $\tau_1^F$ being either 0 or
1\cite{U6/12}.  One can therefore use a model Hamiltonian of the form 
\begin{widetext}
\begin{eqnarray}
 H=&-&\frac{\hbar^2}{2B} \left[ \frac{1}{\beta^4}\frac{\partial}{\partial\beta}\beta^4
\frac{\partial}{\partial\beta}+\frac{1}{\beta^2\sin{3\gamma}}\frac{\partial}{\partial\gamma}\sin{3\gamma}\frac{\partial}{\partial\gamma} 
-\frac{1}{4\beta^4}\sum_\kappa\frac{Q^2_\kappa}{\sin^2{\left(\gamma-\frac{2}{3}\pi\kappa\right)}} \right]
 \nonumber \\ &+&
 u(\beta)+k ~g(\beta)[2\hat \mathcal{L}_B \circ \hat {\mathcal L}_F] + k'~g(\beta) \hat {\mathcal L}_F^2,
  \nonumber \\ &&
\label{Hbohr}
\end{eqnarray}
\end{widetext}
with
\begin{equation}
u(\beta)=0,\beta<\beta_w;~u(\beta)=\infty,\beta \ge \beta_w;
~g(\beta)=\frac{\hbar^2}{2B\beta^2}~,
\end{equation}
where $\hat {\mathcal L}_B$ and $\hat {\mathcal L}_F$ are the
five-dimensional boson and fermion orbital angular momenta, and 
the dot $\circ$ indicates the five-dimensional scalar product. 
This Hamiltonian is the generalization of the Hamiltonian used in the
E(5/4) model. It should be noted that the last term was not
included in E(5/4). 
In that case a single-j was treated and, consequently, a single
$O_F(5)$ representation was involved. For the E(5/12) multi-j case two
$O_F(5)$ representations are present and the last term in
(\ref{Hbohr}) is needed in order to include the energy difference between 
$j=1/2$ and $j=3/2,5/2$ single particle orbits.
The choice of an infinite square well in the $\beta$ 
variable should mimic the actual situation at the critical point of
the transition from the spherical situation (potential with a minimum
in $\beta$=0) to the deformed $\gamma$-instability (potential with a
minimum for a finite value of $\beta$).  In this case, since the
group characterizing the core transition is the Euclidean one in
five dimensions E(5) and the algebra for the fermion space is
U(12), we will name the model as E(5/12). 

With this choice the total wave function can be factorized in the form
\begin{equation}
\Psi~=~F(\beta)~[\Phi(\gamma,\theta_i,\eta_{orb})\otimes \chi(\eta_{spin})]^{(J_{BF})}~,
\end{equation}
where $\eta_{orb}$ and $\eta_{spin}$ represent the particle
coordinates associated with the pseudo-orbital and the pseudo-spin
spaces.  The wave function $\Phi$ is characterized by good quantum
numbers $(\tau_1,\tau_2,\tau_1^B,\tau_1^F,L_{BF},M_L)$ and is given  
by
\begin{widetext}
\begin{eqnarray}
\Phi_{(\tau_1,\tau_2),\tau_1^B,\tau_1^F,L_{BF},M_L}(\gamma,\theta_i,\eta_{orb})~&=&~
\sum_{{\ell}_B,m_{{\ell}_B},{\ell}_F,m_{{\ell}_F}} \left\langle 
\begin{array}{llll}
(\tau_1^B,0)&(\tau_1^F,0)&|&(\tau_1,\tau_2) \\ ~~ {\ell}_B&~~{\ell}_F&|&~~L_{BF}
\end{array}
\right\rangle
\left\langle
\begin{array}{llll}
{\ell}_B&{\ell}_F&|&L_{BF} \\ m_{{\ell}_B}&m_{{\ell}_F}&|&M_L
\end{array}
\right\rangle
\nonumber \\
&&
\nonumber \\
&\times&~\phi_{(\tau_1^B,0),{\ell}_B,m_{{\ell}_B}}(\gamma, \theta_i)~
X_{(\tau_1^F,0),{\ell}_F,m_{{\ell}_F}}(\eta_{orb})~.
\end{eqnarray}
\end{widetext}
The function $X(\eta_{orb})$ describes the five-dimensional
pseudo-orbital part, while $\chi(\eta_{spin})$ is the pseudo-spin wave  
function, which does not contribute to the energy.
The equation for $F(\beta)$ then acquires the familiar form
\begin{equation}
\left[ -\frac{\hbar^2}{2B} \frac{1}{\beta^4}\frac{\partial}{\partial\beta}\beta^4
\frac{\partial}{\partial\beta}+\frac{\hbar^2}{2B}\frac{\Lambda}{\beta^2}
+ u(\beta) \right]F(\beta)~=~EF(\beta)~,
\end{equation}
with
\begin{eqnarray}
&\Lambda&=\tau_1^B(\tau_1^B+3)+k\left[\tau_1(\tau_1+3)+\tau_2(\tau_2+1)\right.
\nonumber \\
&-&\left.
\tau_1^B(\tau_1^B+3)-\tau_1^F(\tau_1^F+3)\right]+k'\tau_1^F(\tau_1^F+3).
\end{eqnarray}

The solutions of this equation, as in the case of the E(5/4) model,
are given in terms of Bessel functions of order
$\nu=\sqrt{\Lambda+9/4}$, in the form 
\begin{eqnarray}
&&F_{\xi,\{(\tau_1,\tau_2),\tau_1^B,\tau_1^F\}}(\beta)~=~c_{\xi,\{(\tau_1,\tau_2),\tau_1^B,\tau_1^F\}}
\nonumber \\
&& \times \beta^{-3/2}
J_\nu(x_{\xi,\{(\tau_1,\tau_2),\tau_1^B,\tau_1^F\}}\beta/\beta_w)~,
\end{eqnarray}
where $c_{\xi,\{(\tau_1,\tau_2),\tau_1^B,\tau_1^F\}}$ is a normalization constant and
$x_{\xi,\{(\tau_1,\tau_2),\tau_1^B,\tau_1^F\}}$ the $\xi$-th zero of $J_{\nu}(z)$.  The corresponding eigenvalues are given as
\begin{equation}
E_{\xi,\{(\tau_1,\tau_2),\tau_1^B,\tau_1^F\}}~=~\frac{\hbar^2}{2B} \left(\frac{
x_{\xi,\{(\tau_1,\tau_2),\tau_1^B,\tau_1^F\}}}{\beta_w} \right)^2.
\end{equation}

\begin{figure}
\begin{center}
\mbox{\epsfig{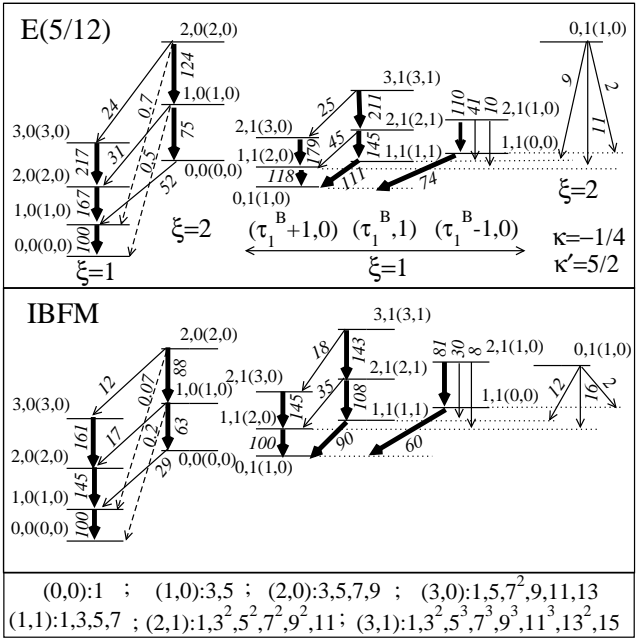}}
\end{center}
\caption{ 
Upper frame:
Energy levels in the odd system at the E(5/12) critical point
(normalized to the energy of the first excited state). Each
(degenerate) 
state is characterized by the $\xi, \tau_1^B,\tau_1^F,(\tau_1,
\tau_2)$ quantum numbers. The values $k=-1/4, k'=5/2$ have been used
in the Hamiltonian (\ref{Hbohr}). The values of (twice) the angular
momenta for each degenerate state can be extracted from the inset in
the lowest panel. Some  
selected values of $B(E2)$'s, corresponding to the transitions between
the highest spins of initial and final multiplets, are given  
in the figure (the B(E2) associated with the lowest 5/2-1/2 transition is 
normalized to 100). 
Lower frame:
Energy levels in the odd system  within the IBFM Hamiltonian (\ref{HBF}) at the critical  point. A number of bosons $N=7$ has been
assumed. The values of $k$ and $k'$ are the same as in the E(5/12) Hamiltonian.
For the BE(2)'s, see the upper frame.
}
\label{fig1}
\end{figure}

The corresponding spectrum 
is given in the upper frame of Fig.~\ref{fig1} for the choice  
of the parameters $k=-1/4$ and $k'=5/2$.   Each state is 
characterized by the $\xi, \tau_1^B,\tau_1^F(\tau_1, \tau_2)$ 
quantum numbers and  arranged in bands according to the counting 
quantum number $\xi$.  The factorization of the pseudo-spin part 
give rise to repeated level couplets. Most of the states are 
therefore degenerate, starting from the first excited 3/2-5/2
doublet. The values of the angular momenta (twice) for each degenerate
state characterized by the $O^{BF}(5)$ boson-fermion $(\tau_1,\tau_2)$ labels  
are given in the lowest inset panel. The lowest band with $\xi=1$ comes from
the coupling of the boson states $(\tau_1^B,0)$ with the fermion
states $(\tau_1^F=0,0)$. It is worth mentioning that all these states
with $\tau_1^F$=0 
have precisely the same energies as the states in the even  
E(5) model. The set of three bands associated to the first excited
$\xi=1$ structure comes from the coupling of $(\tau_1^B,0)$ with
$(\tau_1^F=1,0)$ which produces states $(\tau_1,\tau_2)$ of the full
boson-fermion system labelled as $(\tau_1^B+1,0)$, $(\tau_1^B,1)$ and
$(\tau_1^B-1,0)$\cite{U6/12}. In addition to these $\xi=1$ bands,
excited $\xi=2, 3, \dots$ structures appear.

We can now calculate quadrupole transitions. In this case only the
collective part  
of the quadrupole transition operator has been used, as in
Refs. \cite{E54,cam,caprio}.  The calculation reduces to  
a single numerical integral over the $\beta$ variable. In Fig. \ref{fig1} we
have chosen to represent the B(E2)'s  
between the highest spins of initial and final multiplets (the B(E2)
associated with the lowest 5/2-1/2  
transition is normalized to 100), since most levels are multiply
degenerate. Note that states with different  
$\tau_1^F$ are not connected between them by E2 transitions.

To check the validity of the model we can describe a similar physical
situation  within the framework of the  
Interacting Boson Fermion Model (IBFM)\cite{IVI91}, where the coupling
of a single fermion  to the even-even  
boson core is described by the Hamiltonian
\begin{equation}
H~=~H_{B}+H_{F}+V_{BF}~,
\label{HBF1}
\end{equation}
where the term $V_{BF}$ couples the boson and fermion degrees of freedom. 
We remind that critical point symmetries, as E(5), X(5) or E(5/4), although inspired by 
models with interacting bosons, can only be approximately obtained within these models.
However the comparison with the IBM and IBFM turned out  to be useful, e.g. in the case of the E(5) and E(5/4) symmetries, for a better  
understanding of the characteristics of these models\cite{beta4,odd,odd2}. In these
cases, the Hamiltonian used  
for the bosonic part was a parametrized one, that produces a
transition between spherical and  
$\gamma$-unstable shapes, of the form
\begin{equation}
H_B= x \hat n_d - 
((1-x)/N) 
~\hat Q_B \cdot \hat Q_B ~,
\label{HQQ}
\end{equation}
where the operators in the Hamiltonian are given by
$\hat n_d  =  \sum_\mu d^\dag _\mu d_\mu$
and  
$\hat Q_B  = (s^\dag \times \tilde d + d^\dag \times \tilde s)^{(2)}$,
and $N$ is the total number of bosons.  This Hamiltonian can be recast into the form
\begin{equation}
H_B= xC_1(U^B(5))-\frac{1-x}{2N}\left[ C_2(O^B(6))-C_2(O^B(5))\right],
\label{HB_casimir}
\end{equation}
where $C_1(U^B(5))$ is the linear Casimir operator of the $U^B(5)$
algebra, while $C_2(O^B(6))$ and  
$C_2(O^B(5))$ are the quadratic Casimir operators of the $O^B(6)$ and
$O^B(5)$ algebras.  By using the intrinsic frame formalism (see for instance
\cite{jolie1}), one can show that the critical point occurs at
$x_{crit}=\frac{4N-8}{5N-8}$. 

In analogy with this boson Hamiltonian, which has been extensively
studied is Refs. \cite{iach5,rowe,beta4},  
one could choose\cite{athens} in our case a boson-fermion Hamiltonian of the form
\begin{eqnarray}
H&=& x \left( \hat n_d+ \hat n_{3/2}+ \hat n_{5/2} \right) -
\frac{1-x}{N} \hat Q_{BF}\cdot \hat Q_{BF} ~ \nonumber  \\
   &=&xC_1(U^{BF}(5)) \nonumber \\
&-& \frac{1-x}{2N}\left[ C_2(O^{BF}(6))-C_2(O^{BF}(5))\right],
\label{HBF_casimir}
\end{eqnarray}
which implies a certain choice of the total quadrupole operator and
quasiparticle energies \cite{IVI91}.  
In particular, it is worth noting that the $j=3/2$ and $5/2$ orbits are
degenerate and higher in energy than the $j=1/2$ orbit. Such a Hamiltonian has been used thoroughly in the literature
and, in particular, in a previous study of shape phase transition in
odd-mass nuclei using a supersymmetric approach\cite{jolie2}. It has the advantage that one recovers, in the extreme cases, the Bose-Fermi
symmetry\cite{IVI91} associated with $O^{BF}(6) \otimes U^F_s(2)$ (for
$x$=0) and the vibrational $U^{BF}(5)\otimes U^F_s(2)$ (for $x$=1). 
On the other hand, such an IBFM Hamiltonian only approximately preserves \cite{athens} the 
$\tau^B_1$ and $\tau^F_1$ quantum numbers, while those are good quantum numbers for E(5/12). 
Although it will be possible to establish anyway a correspondence between E(5/12) and IBFM states, selection rules for electromagnetic and one-particle transfer transitions will be different.

\begin{figure}
\begin{center}
\mbox{\epsfig{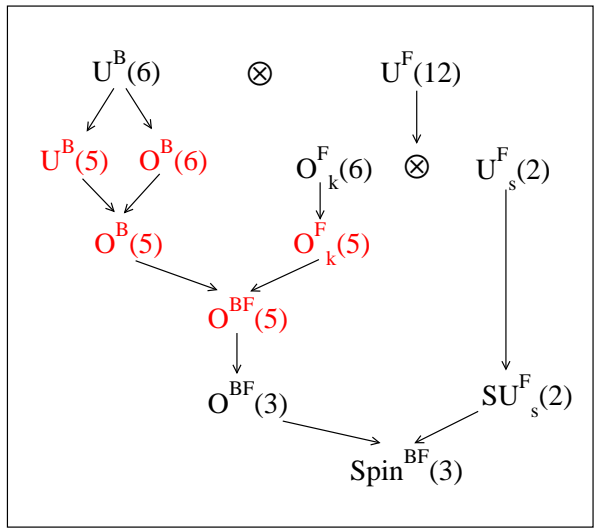}}
\end{center}
\caption{The lattice of algebras involved in the IBFM Hamiltonian (\ref{HBF}).
}
\label{fig2}
\end{figure}

For a better comparison with the E(5/12) results we have therefore preferred to use here
at the critical point 
a boson-fermion Hamiltonian of the form
\begin{eqnarray}
H~&=&~ xC_1(U^B(5))-\frac{1-x}{2N}\left[ C_2(O^B(6))-C_2(O^B(5))\right]\nonumber \\
 &+& \frac{k}{2N}\left[ C_2(O^{BF}(5))-C_2(O^B(5))-C_2(O^F(5))\right]\nonumber \\
 &+&  \frac{k'}{2N}~C_2(O^F(5))~, 
\label{HBF}
\end{eqnarray}
where $C_2(O^{F}(5))$,   
$C_2(O^{BF}(5))$ are the quadratic Casimir operators of the $O^F(5)$ and $O^{BF}(5)$ algebras. This Hamiltonian is designed to mimic as much as possible the corresponding
Hamiltonian in E(5/12), since the terms 
$2\hat \mathcal{L}_B \circ \hat {\mathcal L}_F$ and  $ \hat {\mathcal L}_F^2$ in (\ref{Hbohr}) can be written in terms of Casimir operators as
$2\hat \mathcal{L}_B \circ \hat {\mathcal L}_F=(1/2)[
C_2(O^{BF}(5))-C_2(O^B(5))-C_2(O^F(5))]$ and  $ \hat {\mathcal L}_F^2=
(1/2)C_2(O^F(5))$.
This 
boson-fermion Hamiltonian preserves by construction the $\tau^B_1$ and $\tau^F_1$ quantum numbers and all states are therefore characterized by the same quantum numbers and degeneracies as in the E(5/12) case.   For a better clarification, we show in Fig.~\ref{fig2} the lattice of algebras relevant for our problem.

The corresponding
spectrum is given in the lower frame of  Fig.~\ref{fig1}. A number N=7 of bosons is assumed as a typical value.  Other choices
of N do not alter the qualitative behaviour of our results. With the
choice  N=7 the critical point occurs at
$x_{crit}=\frac{4N-8}{5N-8}=\frac{20}{27}\approx 0.74$.  
In the Hamiltonian the values for $k$ and $k'$ have been assumed to be equal to the corresponding values in the E(5/12) case presented in the upper frame.

The spectrum has a structure that clearly resembles the one obtained
within the E(5/12) model (upper frame).   
Energies of selected states in the two models are compared in Table~{\ref{levels}}.  
As noted before, in the E(5/12) model all states with $\tau_1^F$=0 (i.e. with the odd particle in the j=1/2 state) have precisely the same energies as the states in the even E(5) model.  Similarly, in the IBFM these states
have the same energy of the corresponding states in the boson Hamiltonian (\ref{HQQ}).
The agreement for the energies of these states in E(5/12) and IBFM is therefore that already discused in ref.\cite{beta4} for the bosonic part. Excited bands 
with $\tau_1^F$=1 (i.e. with the odd particle in the j=3/2,5/2 states), on the other hand, seem  to be somewhat compressed in E(5/12) with respect to the IBFM.  

The same kind of agreement is also present in the
electromagnetic transitions.  
Consistently with the choice presented in the E(5/12), 
we assume in the IBFM
as E2 transition operator just the boson (collective) 
quadrupole operator, i.e. $\hat T^{E2}~=~ \hat Q_B$.
The order of magnitude of the different B(E2) values are reflected in different  
thickness and styles of the arrows shown in Fig.~\ref{fig1}.   Given the choice of the IBFM Hamiltonian (\ref{HBF}) the same selection rules valid in the E(5/12) model also apply to the ($\tau_1^B,\tau_1^F, \tau_1,\tau_2)$ quantum numbers of initial and final states in the IBFM approach. The B(E2) values of selected transitions are explicitly given in the figure.  All B(E2) values are normalized to 100 for the first 5/2 - 1/2 transition.
Similarly to what happens for the energies  of the levels, the overall
behaviour is very similar to that of the E(5/12).  

\begin{table}[h]
\begin{center}
\begin{tabular}
{|c|c|c|}
\tableline
&&\\
{\bf Level}&{\bf Energy}
&{\bf Energy} \\
$~~\xi,\tau_1^B,\tau_1^F (\tau_1, \tau_2)~~$&~~{\bf E(5/12)}~~ & 
  ~~{\bf IBFM}~~\\
  &&\\
\tableline
&&\\
1,1,0 (1,0) & 1.00
 &  1.00 \\
1,2,0 (2,0) & 2.20
 &  2.30\\
2,0,0 (0,0) & 3.03
& 2.86  \\
1,0,1 (1,0) & 2.20
 & 2.69 \\
1,1,1 (1,1) & 3.00
& 3.82 
 \\
 && \\
\tableline
\end{tabular}
\end{center}
\begin{center}
\caption{
Energies of selected states in  E(5/12) and IBFM (normalized to the
first excited multiplet).  The values $k=-1/4, k'=5/2$ have been used in both Hamiltonians
(\ref{Hbohr}) and (\ref{HBF}). In the IBFM calculation,  a number N=7 of bosons has been used.
 }
\label{levels}
\end{center}
\end{table}

In this letter we have presented an analytic model (E(5/12)) for the
critical point of the phase transition from spherical to
$\gamma$-unstable shapes in odd nuclei, i.e. a mixture of fermion and
boson degrees of freedom. In the model the fermion is allowed to occupy
many single particle orbitals.  Spectrum and intensities of
electromagnetic transitions have been compared with those obtained
within  
the Interacting Boson Fermion Model showing comparable behaviours.  We
think that our model,  which includes the more realistic many j-orbits
case compared with the E(5/4) model, should offer more chances for the
occurrence of critical point symmetries in odd nuclei, in addition to
those already found in the case of even nuclei.  

This work has been partially supported by the Spanish Ministerio 
de Educaci\'on y Ciencia 
and by the European regional development fund (FEDER) under 
project number FIS2005-01105, and by INFN. Andrea Vitturi acknowledges 
financial support from SEUI (Spanish Ministerio de Educaci\'on y 
Ciencia) for a sabbatical year at University of Sevilla.

\end{document}